\documentclass[prd,aps,nofootinbib,floatfix,11pt]{revtex4}

\usepackage{amsmath,graphicx,epsfig,amssymb,dsfont,mathtools}
\usepackage[usenames]{color}
\usepackage{ulem} 
\usepackage{bigstrut}
\usepackage{slashed}
\usepackage{multirow}
\usepackage{subfigure}

\allowdisplaybreaks

\begin{document}

\title{Weak decays of singly heavy baryons: the $1/2\to3/2$ case}
 
\author{
Fu-Wei Zhang$^{1}$, 
Zhen-Xing Zhao$^{1,2}$~\thanks{Email: zhaozx19@imu.edu.cn (Corresponding author)}
}
\affiliation{
$^{1}$ School of Physical Science and Technology, Inner Mongolia University, Hohhot 010021, China\\
 $^{2}$ Research Center for Quantum Physics and Technologies, Inner Mongolia University, Hohhot 010021, China
}
 
\begin{abstract}
This work is devoted to investigating the $1/2\to3/2$ weak decays
of singly heavy baryons. Due to the orthogonality between the spin
wavefunctions of antitriplet baryons and spin-3/2 baryons, the weak
decay amplitude for such processes vanishes at the leading order of QCD. Consequently,
this study exclusively examines the weak decays of $\Omega_{Q}$.
Using the light-front approach under the three-quark picture, we first
extract the relevant form factors, and then apply them to investigate
corresponding semileptonic and nonleptonic decays. Finally, we compare
our phenomenological predictions with existing results in the literature.
Our findings are expected to be helpful in experimentally
establishing these decay channels.
\end{abstract}

\maketitle

\section{Introduction}

Recently, the LHCb collaboration reported the first observation of
CP violation (CPV) in baryon decays through the channel $\Lambda_{b}^{0}\to pK^{-}\pi^{+}\pi^{-}$
\cite{LHCb:2025ray}. This is another significant progress in the
field of particle physics in recent years. It is also worth noting
that, Refs. \cite{Yu:2025ekh,Wang:2024oyi} predicted that
the CPV for $\Lambda_{b}^{0}\to R(p\pi^{+}\pi^{-})K^{-}$
in the mass region $m_{p\pi^{+}\pi^{-}}<2.7\,\mathrm{GeV}/c^{2}$
is $(5.6\text{-}5.9)\%$, which shows excellent agreement with the
LHCb measurement result: ${\cal A}_{CP}(\Lambda_{b}^{0}\to R(p\pi^{+}\pi^{-})K^{-})=(5.4\pm0.9\pm0.1)\%$.
The discovery of baryon CPV is of great significance to explain the
origin and evolution of the universe, and it also opens a new path
to search for new physics beyond the Standard Model.

It can be expected that this discovery will lead to another Renaissance
in the field of heavy flavor baryons. In particular, one might expect
CPV to be found in other baryon decays, which requires us to have
a sufficient understanding of the decay modes of the heavy flavor
baryons. In Ref. \cite{Zhao:2018zcb}, we performed a comprehensive
study on the weak decays of singly heavy baryons for the $1/2\to1/2$
case. As a continuation, in this work we intend to study the $1/2\to3/2$
case. Numerous theoretical studies have investigated the $1/2\to3/2$
transitions of singly heavy baryons \cite{Ivanov:1996fj,Cheng:1996cs,Ebert:2006rp,Ke:2017eqo,Aliev:2022gxi,Wang:2022zja,Liu:2023dvg,Lu:2023rmq,Zeng:2024yiv,Shi:2025ocl,Pervin:2006ie},
however, most of them are typically focused on some specific processes.
Our present work aims to provide a comprehensive investigation on
these processes including multiple previously unexplored channels. 

\begin{figure}
\includegraphics[scale=0.7]{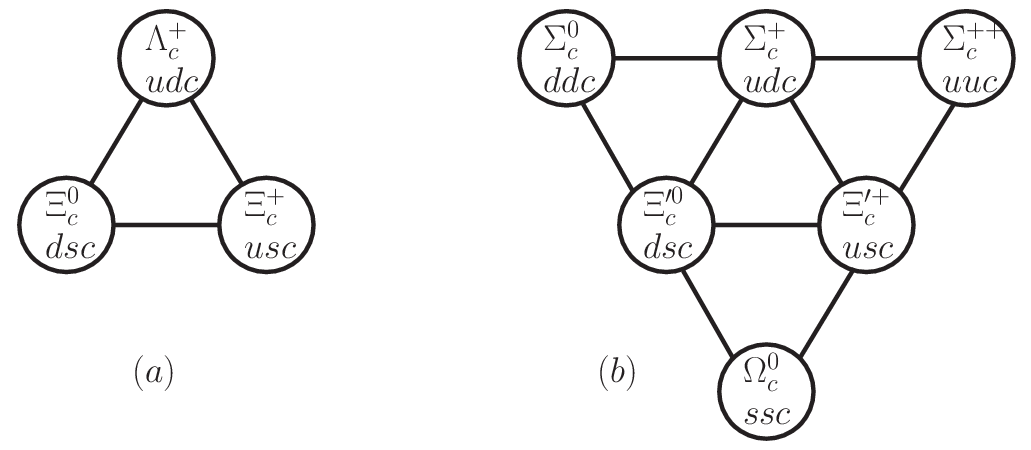}
\caption{
SU(3) flavor symmtry arranges singly heavy baryons into an antitriplet $\bar{\boldsymbol{3}}$ and a sextet $\boldsymbol{6}$.
}
\label{fig:singly_heavy}
\end{figure}

A singly heavy baryon consists of one heavy and two light valence
quarks. SU(3) flavor symmtry arranges them into a set of antitriplet
states $\bar{\boldsymbol{3}}$ and a set of sextet states $\boldsymbol{6}$,
as can be seen from Fig. \ref{fig:singly_heavy}. Specifically, for
charmed baryons, $\bar{\boldsymbol{3}}$ is composed of $\Lambda_{c}^{+}$
and $\Xi_{c}^{+,0}$, while $\boldsymbol{6}$ is composed of $\Sigma_{c}^{++,+,0}$,
$\Xi_{c}^{\prime+,\prime0}$ and $\Omega_{c}^{0}$; for bottom baryons,
$\bar{\boldsymbol{3}}$ is composed of $\Lambda_{b}^{0}$ and $\Xi_{b}^{0,-}$,
while $\boldsymbol{6}$ is composed of $\Sigma_{b}^{+,0,-}$, $\Xi_{b}^{\prime0,\prime-}$
and $\Omega_{b}^{-}$. Only the antitriplet heavy baryons and the
sextet $\Omega_{Q}$ predominantly undergo weak decays, while all
other baryons decay primarily via strong or electromagnetic interactions.
Moreover, at the leading order of QCD, an antitriplet baryon cannot
decay into an S-wave $3/2^{+}$ baryon. This is because their spin wavefunctions are orthogonal to each other. Therefore, in this work, we
will only focus on the $1/2\to3/2$ decay processes of $\Omega_{Q}$
baryons:
\begin{align}
\Omega_{c}^{0}(css) & \rightarrow\Xi^{*-}(dss)/\Omega^{-}(sss),\nonumber \\
\Omega_{b}^{-}(bss) & \rightarrow\Xi^{*0}(uss)/\Omega_{c}^{*0}(css).
\end{align}

We will employ the light-front quark model (LFQM) to study the decay
dynamics, which has been extensively applied in investigating the
decay properties of mesons \cite{Cheng:2004yj,Ke:2009ed,Ke:2010htz,Cheng:2009ms,Lu:2007sg,Enqi:2025vdp,Wang:2007sxa,Wang:2008xt,Wang:2008ci,Wang:2009mi,Chen:2009qk,Li:2010bb,Shi:2016gqt,Cheng:2003sm,Shi:2023qnw}
and baryons \cite{Zhao:2018zcb,Xing:2018lre,Zhao:2018mrg,Zhao:2022vfr,Zhao:2023yuk,Xing:2024okl,Xing:2023jnr,Wang:2022ias,Liu:2022mxv,Liu:2023zvh,Zhu:2018jet,Ke:2017eqo,Wang:2017mqp}.
In the baryon sector, there exist two distinct pictures depending
on how the two spectator quarks are treated. Taking the two spectator
quarks as a whole, the picture is called the ``diquark picture";
Treating the two spectator quarks independently, this picture is called
the ``three-quark picture". Obviously, the former picture
is simpler, but it will inevitably introduce more uncertainties. In
Ref. \cite{Zhao:2023yuk}, we reconstructed the three-quark picture
from bottom to top, and pointed out the crucial role of Lorentz boost,
paving the way for further applications of the quark model.

The rest of the paper is organized as follows. In Sec. II, we will
outline the framework of the light-front quark model in the three-quark
picture and present the corresponding overlap factors. Our main results,
including the calculated form factors and predictions for both semileptonic
and nonleptonic decay widths, are detailed in Sec. III; We also compare
our results with other existing theoretical predictions there. In
the last section, a brief summary will be given.

\section{Theoretical Framework}

\subsection{The light-front approach}

The baryon state in the light-front approach is expressed as
\begin{align}
|\mathcal{B}(P,S,S_{z})\rangle & =\int\{d^{3}\tilde{p_{1}}\}\{d^{3}\tilde{p_{2}}\}\{d^{3}\tilde{p_{3}}\}2(2\pi)^{3}\delta^{3}(\tilde{P}-\tilde{p}_{1}-\tilde{p}_{2}-\tilde{p}_{3})\frac{1}{\sqrt{P^{+}}}\nonumber \\
 & \times\sum_{\lambda_{1},\lambda_{2},\lambda_{3}}\Psi^{SS_{z}}(\tilde{p}_{1},\tilde{p}_{2},\tilde{p}_{3},\lambda_{1},\lambda_{2},\lambda_{3})C^{ijk}|q_{1}^{i}(p_{1},\lambda_{1})q_{2}^{j}(p_{2},\lambda_{2})q_{3}^{k}(p_{3},\lambda_{3})\rangle,
\end{align}
where $p_{i}$ ($\lambda_{i}$) denotes the momentum (helicity) of
the $i$-th quark, $C^{ijk}=\epsilon^{ijk}/\sqrt{6}$ represents the
totally antisymmetric color wavefunction, and $\Psi^{SS_{z}}$ incorporates
the momentum and flavor-spin wavefunctions.

In the light-front approach, the momentum is decomposed into its light-front
form $p_{i}=(p_{i}^{-},p_{i}^{+},p_{i\perp})$ with $p_{i}^{\pm}=p_{i}^{0}\pm p_{i}^{3}$
and $p_{i\perp}=(p_{i}^{1},p_{i}^{2})$, the quarks are treated to
be on the mass shell 
\begin{equation}
p_{i}^{-}=\frac{m_{i}^{2}+p_{i\perp}^{2}}{p_{i}^{+}},
\end{equation}
and 
\begin{equation}
\tilde{p_{i}}=(p_{i}^{+},p_{i\perp}),\quad\{d^{3}\tilde{p_{i}}\}=\frac{dp_{i}^{+}d^{2}p_{i\perp}}{2(2\pi)^{3}}.
\end{equation}

The intrinsic variables $(x_{i},k_{i\perp})$ are given as follows
\begin{equation}
p_{i}^{+}=x_{i}P^{+},\quad p_{i\perp}=x_{i}P_{\perp}+k_{i\perp},\quad\sum_{i=1}^{3}x_{i}=1,\quad\sum_{i=1}^{3}k_{i\perp}=0,
\end{equation}
where $x_{i}$ is the light-front momentum fraction, which is constrained
by $0\leq x_{i}\leq1$. After defining $\bar{P}=p_{1}+p_{2}+p_{3}$
and $M_{0}^{2}=\bar{P}^{2}$, one can check that
\begin{equation}
M_{0}^{2}=\frac{k_{1\perp}^{2}+m_{1}^{2}}{x_{1}}+\frac{k_{2\perp}^{2}+m_{2}^{2}}{x_{2}}+\frac{k_{3\perp}^{2}+m_{3}^{2}}{x_{3}}.
\end{equation}

The internal momenta are defined as
\begin{equation}
k_{i}=(k_{i}^{-},k_{i}^{+},k_{i\perp})=(e_{i}-k_{iz},e_{i}+k_{iz},k_{i\perp}),
\end{equation}
then it is easy to arrive at
\begin{align}
e_{i} & =\frac{x_{i}M_{0}}{2}+\frac{m_{i}^{2}+k_{i\perp}^{2}}{2x_{i}M_{0}},\nonumber \\
k_{iz} & =\frac{x_{i}M_{0}}{2}-\frac{m_{i}^{2}+k_{i\perp}^{2}}{2x_{i}M_{0}},
\end{align}
where $e_{i}$ is the energy of the $i$-th quark in the rest frame of
$\bar{P}$.

In the three-quark picture of the light-front approach, three typical
wavefunctions are given as follows: 
\begin{itemize}
\item $\Lambda_{Q}$-type, where the two light quarks are treated to form
a $0^{+}$ diquak, 
\begin{align}
\Psi_{0}^{S=\frac{1}{2},S_{z}}(\tilde{p_{i}},\lambda_{i}) & =A_{0}\bar{u}(p_{3},\lambda_{3})(\bar{\slashed{P}}+M_{0})(-\gamma_{5})C\bar{u}^{T}(p_{2},\lambda_{2})\nonumber \\
 & \times\bar{u}(p_{1},\lambda_{1})u(\bar{P},S_{z})\Phi(x_{i},k_{i\perp}),
\end{align}
\item $\Sigma_{Q}$-type, where the two light quarks are treated to form
a $1^{+}$ diquak, 
\begin{align}
\Psi_{1}^{S=\frac{1}{2},S_{z}}(\tilde{p_{i}},\lambda_{i}) & =A_{1}\bar{u}(p_{3},\lambda_{3})(\bar{\slashed{P}}+M_{0})(\gamma^{\mu}-\upsilon^{\mu})C\bar{u}^{T}(p_{2},\lambda_{2})\nonumber \\
 & \times\bar{u}(p_{1},\lambda_{1})(\frac{1}{\sqrt{3}}\gamma_{\mu}\gamma_{5})u(\bar{P},S_{z})\Phi(x_{i},k_{i\perp}),
\end{align}
\item $\Sigma_{Q}^{*}$-type, where the two light quarks are also treated
to form a $1^{+}$ diquark, and the total spin is $3/2$, 
\begin{align}
\Psi_{1}^{S=\frac{3}{2},S_{z}}(\tilde{p_{i}},\lambda_{i}) & =A_{1}^{\prime}\bar{u}(p_{3},\lambda_{3})(\bar{\slashed{P}}+M_{0})(\gamma^{\mu}-\upsilon^{\mu})C\bar{u}^{T}(p_{2},\lambda_{2})\nonumber \\
 & \times\bar{u}(p_{1},\lambda_{1})u_{\mu}(\bar{P},S_{z})\Phi(x_{i},k_{i\perp}),
\end{align}
\end{itemize}
where $v=\bar{P}/M_{0}$, $\Phi$ is the momentum wavefunction, and
the normalization factors
\begin{equation}
A_{0}=A_{1}=A_{1}^{\prime}=\frac{1}{4\sqrt{M_{0}^{3}(e_{1}+m_{1})(e_{2}+m_{2})(e_{3}+m_{3})}}.
\end{equation}

We adopt the following momentum wavefunction:
\begin{equation}
\Phi(x_{i},k_{i\perp})=\sqrt{\frac{e_{1}e_{2}e_{3}}{x_{1}x_{2}x_{3}M_{0}}}\varphi\left(\vec{k}_{1},\beta_{1}\right)\varphi\left(\frac{\vec{k}_{2}-\vec{k}_{3}}{2},\beta_{23}\right),\label{eq:T_type}
\end{equation}
where $\varphi(\vec{k},\beta)\equiv4\left(\frac{\pi}{\beta^{2}}\right)^{3/4}\exp\left(-\frac{\vec{k}^{2}}{2\beta^{2}}\right)$
with $\vec{k}\equiv(k_{\perp},k_{z})$, and $\beta_{1}$ and $\beta_{23}$
are the so-called shape parameters that characterize the momentum
distribution inside the baryon.

\subsection{Form factors}

The weak decay matrix element can be parameterized in terms of form
factors as 
\begin{align}
 & \langle{\cal B}_{f}(P^{\prime},S^{\prime}=\frac{3}{2},S_{z}^{\prime})|\bar{q}\gamma^{\mu}Q|{\cal B}_{i}(P,S=\frac{1}{2},S_{z})\rangle\nonumber \\
= & \bar{u}_{\alpha}(P^{\prime},S_{z}^{\prime})[\gamma^{\mu}\frac{P^{\alpha}}{M}f_{1}(q^{2})+\frac{P^{\alpha}P^{\mu}}{M^{2}}f_{2}(q^{2})+\frac{P^{\alpha}P^{\prime\mu}}{MM^{\prime}}f_{3}(q^{2})+g^{\mu\alpha}f_{4}(q^{2})]\gamma_{5}u(P,S_{z}),\label{eq:fi}\\
 & \langle{\cal B}_{f}(P^{\prime},S^{\prime}=\frac{3}{2},S_{z}^{\prime})|\bar{q}\gamma^{\mu}\gamma_{5}Q|{\cal B}_{i}(P,S=\frac{1}{2},S_{z})\rangle\nonumber \\
= & \bar{u}_{\alpha}(P^{\prime},S_{z}^{\prime})[\gamma^{\mu}\frac{P^{\alpha}}{M}g_{1}(q^{2})+\frac{P^{\alpha}P^{\mu}}{M^{2}}g_{2}(q^{2})+\frac{P^{\alpha}P^{\prime\mu}}{MM^{\prime}}g_{3}(q^{2})+g^{\mu\alpha}g_{4}(q^{2})]u(P,S_{z}),\label{eq:gi}
\end{align}
where $q=P-P^{\prime}$, $f_{i}$ and $g_{i}$ are respectively the
form factors of the vector and axial-vector current, and $M$ and
$M^{\prime}$ are the masses of the initial and final baryons. The
matrix element can also be calculated in LFQM (Here we take the $\Omega_{c}^{0}(css)\rightarrow\Xi^{*-}(dss)$ transition as an example)
\begin{align}
\langle\Xi^{*-}|\bar{d}\gamma^{\mu}(1-\gamma_{5})c|\Omega_{c}^{0}\rangle & =\int\{d^{3}\tilde{p_{2}}\}\{d^{3}\tilde{p_{3}}\}\frac{A_{1}^{\prime}A_{1}}{\sqrt{p_{1}^{\prime+}p_{1}^{+}P^{\prime+}P^{+}}}\Phi^{\prime*}(x_{i}^{\prime},k_{i\perp}^{\prime})\Phi(x_{i},k_{i\perp})\times\frac{1}{\sqrt{2}}\times\frac{1}{\sqrt{2}}\times2\nonumber \\
 & \times\text{Tr}[C(\gamma^{\sigma}-v^{\prime\sigma})(\bar{\slashed P}^{\prime}+M_{0}^{\prime})(\slashed p_{3}+m_{3})(\bar{\slashed P}+M_{0})(\gamma^{\rho}-v^{\rho})C(\slashed p_{2}+m_{2})^{T}]\nonumber \\
 & \times\bar{u}_{\sigma}(\bar{P}^{\prime},S_{z}^{\prime})(\slashed p_{1}^{\prime}+m_{1}^{\prime})\gamma^{\mu}(1-\gamma_{5})(\slashed p_{1}+m_{1})(\frac{1}{\sqrt{3}}\gamma_{\rho}\gamma_{5})u(\bar{P},S_{z}),\label{eq:me_LFQM}
\end{align}
where $v^{\rho}=\bar{P}^{\rho}/M_{0}$, $v^{\prime\sigma}=\bar{P}^{\prime\sigma}/M_{0}^{\prime}$,
and the overlap factor $1/\sqrt{2}\times1/\sqrt{2}\times2$ consists
of three parts: the normalization of the initial state,
the normalization of the final state, and $2$ equivalent contractions.
The (vector current) form factors can be extracted by the following steps: (1) Multiply
Eq. (\ref{eq:fi}) by $\bar{u}(P,S_{z})(\Gamma_{5}^{\mu\beta})_{i}u_{\beta}(P^{\prime},S_{z}^{\prime})$
with $(\Gamma_{5}^{\mu\beta})_{i}=\{\gamma^{\mu}P^{\beta},P^{\prime\mu}P^{\beta},P^{\mu}P^{\beta},g^{\mu\beta}\}\gamma_{5}$
to obtain a set of expressions; (2) Do the same thing for the vector-current
part of Eq. (\ref{eq:me_LFQM}) to obtain another set of expressions,
meanwhile take the approximation $P\rightarrow\bar{P}$ and $P^{\prime}\rightarrow\bar{P}^{\prime}$
in the integral; (3) Extract the form factors by equating the two
sets of expressions. 

\subsection{Overlap factors}

One can see from Eq. (\ref{eq:me_LFQM}) that the overlap factor appears
in the expression of the transition matrix element. In the diquark
picture, its definition is vague. Only in the three quark picture
can this factor be clearly defined -- it is actually the inner product
of the flavor wavefunctions of the initial and final baryons. It consists
of the normalization factors of initial and final states, and the
contraction factor. All the overlap factors are collected in Table
\ref{Tab:overlap_factors}. 

\begin{table}
\caption{Overlap factors in flavor space.}
\label{Tab:overlap_factors} %
\begin{tabular}{c|c|c|c}
\hline 
Transition & Overlap factor & Transition & Overlap factor\tabularnewline
\hline 
$\Omega_{c}^{0}(css)\rightarrow\Xi^{*-}(dss)$  & $\frac{1}{\sqrt{2}}\times\frac{1}{\sqrt{2}}\times2$  & $\Omega_{c}^{0}(css)\rightarrow\Omega^{-}(sss)$  & $\frac{1}{\sqrt{2}}\times\frac{1}{\sqrt{6}}\times6$ \tabularnewline
\hline 
$\Omega_{b}^{-}(bss)\rightarrow\Xi^{*0}(uss)$  & $\frac{1}{\sqrt{2}}\times\frac{1}{\sqrt{2}}\times2$ & $\Omega_{b}^{-}(bss)\rightarrow\Omega_{c}^{*0}(css)$  & $\frac{1}{\sqrt{2}}\times\frac{1}{\sqrt{2}}\times2$\tabularnewline
\hline 
\end{tabular}
\end{table}

\section{Numerical results and phenomenological applications}

In this section, we first present the numerical results of the form
factors. Subsequently, these form factors will be applied to predict
some phenomenological observables, including the decay widths of corresponding
semileptonic and nonleptonic decays. For the latter case, we are constrained
to consider only the factorizable W-emission diagram. For example, for $\Omega_{c}^{0}\to\Omega^{-}\pi^{+}$, only the factorizable diagram contributes at the leading order of QCD.
However, some other nonleptonic decays may also receive contributions
from non-factorizable diagrams, for which, our predictions here can only
be considered as rough estimates. However, considering that nonleptonic
decays are of practical significance for experimental findings, we
still believe that our estimates are valuable. 

\subsection{Inputs}

The constituent quark masses and shape parameters are the main model
parameters of LFQM. In this work, we adopt the following constituent
quark masses (in units of GeV):
\begin{align}
m_{u} & =m_{d}=0.25,\;m_{s}=0.37,\;m_{c}=1.4,\;m_{b}=4.8.
\end{align}
For the selection of the shape parameters, we mainly refer to Ref.
\cite{Zhao:2023yuk}, where we obtained (in units of GeV):
\begin{equation}
\beta_{\{ud\}}=0.28,\;\beta_{\{cc\}}=0.40,\;\beta_{c\{ud\}}=0.49,\;\beta_{b\{ud\}}=0.66.
\end{equation}
It is reasonable that: 
\begin{itemize}
\item $\beta_{\{ss\}}$ should be slightly larger than $\beta_{\{ud\}}$
and much smaller than $\beta_{\{cc\}}$;
\item Given that the shape parameters reflect the size of the baryon, we
can use $\beta_{\{ss\}}$ to determine $\beta_{s\{ss\}}$;
\item $\beta_{q\{ss\}}$ with $q=u/d$ should be larger than $\beta_{\{ud\}}$
and smaller than $\beta_{s\{ss\}}$;
\item $\beta_{c\{ss\}}$ and $\beta_{b\{ss\}}$ should be slightly larger
than $\beta_{c\{ud\}}$ and $\beta_{b\{ud\}}$, respectively.
\end{itemize}
In the end, we choose the following shape parameters (in units of
GeV) for this work:
\begin{equation}
\beta_{\{ss\}}=0.30,\;\beta_{q\{ss\}}=0.33,\;\beta_{s\{ss\}}=0.35,\;\beta_{c\{ss\}}=0.51,\;\beta_{b\{ss\}}=0.68.
\end{equation}
These shape parameters carry approximately 10\% uncertainties, which will be used for error estimation in the following text.

The masses (in units of GeV) of initial and final baryons are listed
as follows \cite{ParticleDataGroup:2024cfk}:
\begin{align}
m_{\Omega_{c}} & =2.695,\;m_{\Omega_{b}}=6.046,\;m_{\Xi^{*-}}=1.534,\;m_{\Omega^{-}}=1.672,\;m_{\Omega_{c}^{*}}=2.766.
\end{align}
The lifetimes of initial baryons are \cite{ParticleDataGroup:2024cfk}
\begin{equation}
\tau_{\Omega_{b}}=(1.64\pm0.16)\times10^{-12}\,\text{s},\quad\tau_{\Omega_{c}}=(2.73\pm0.12)\times10^{-13}\,\text{s}.
\end{equation}
The Fermi constant and CKM matrix elements are \cite{ParticleDataGroup:2024cfk}
\begin{align}
 & G_{F}=1.166\times10^{-5}\mathrm{GeV}^{-2},\nonumber \\
 & |V_{ud}|=0.974,\quad|V_{us}|=0.225,\quad|V_{ub}|=0.00373,\nonumber \\
 & |V_{cd}|=0.225,\quad|V_{cs}|=0.973,\quad|V_{cb}|=0.0418.
\end{align}
The masses and decay constants (in units of MeV) of mesons in nonleptonic
decays are 
\begin{align}
 & m_{\pi}=140,\;m_{\rho}=775,\;m_{a_{1}}=1230,\;m_{K}=494,\;m_{K^{*}}=892,\nonumber \\
 & m_{D}=1870,\;m_{D^{*}}=2007,\;m_{D_{s}}=1968,\;m_{D_{s}^{*}}=2107,\\
 & f_{\pi}=130.4,\;f_{\rho}=216,\;f_{a_{1}}=238,\;f_{K}=160,\;f_{K^{*}}=210,\nonumber \\
 & f_{D}=207.4,\;f_{D^{*}}=220,\;f_{D_{s}}=247.2,\;f_{D_{s}^{*}}=247.2,\nonumber 
\end{align}
which can be found in Refs. \cite{Carrasco:2014poa,Cheng:2003sm,Shi:2016gqt,Zhao:2018mrg}.
The Wilson coefficient $a_{1}(\mu)\equiv C_{1}(\mu)+C_{2}(\mu)/3$
is taken as $a_{1}(\mu_{b})=1.03$ for the bottom decay and $a_{1}(\mu_{c})=1.10$
for the charm decay \cite{Buras:1998raa}. 

\subsection{Form factors}

The calculated form factors at $q^{2}=0$ are given in Table \ref{Tab:ff0}.
Next we adopt the single-pole assumption 
\begin{equation}
F(q^{2})=\frac{F(0)}{1-q^{2}/m_{\mathrm{pole}}^{2}}
\end{equation}
to access the $q^{2}$ dependence, where the masses of $D$, $D_{s}$,
$B$ and $B_{c}$ are taken as $m_{\mathrm{pole}}$ for $c\rightarrow d/s$
and $b\rightarrow u/c$, respectively. For the validity of this assumption,
readers can refer to Ref. \cite{Shi:2016gqt}.

\begin{table}
\caption{Our predicted form factors at $q^{2}=0$, where the uncertainties
from the shape parameters are considered. }
\label{Tab:ff0} %
\resizebox{\textwidth}{!}{
\begin{tabular}{c|c|c|c|c|c|c|c|c}
\hline 
Transition & $f_{1}(0)$ & $f_{2}(0)$ & $f_{3}(0)$ & $f_{4}(0)$ & $g_{1}(0)$ & $g_{2}(0)$ & $g_{3}(0)$ & $g_{4}(0)$\tabularnewline
\hline 
$\Omega_{c}^{0}\rightarrow\Xi^{*-}$ & $-0.591\pm0.067$ & $0.056\pm0.039$ & $0.490\pm0.009$ & $-1.230\pm0.122$ & $-0.303\pm0.193$ & $0.288\pm0.072$ & $-0.313\pm0.138$ & $0.615\pm0.078$\tabularnewline
\hline 
$\Omega_{c}^{0}\rightarrow\Omega^{-}$ & $-1.283\pm0.124$ & $0.164\pm0.120$ & $1.022\pm0.037$ & $-2.622\pm0.218$ & $-2.198\pm0.840$ & $0.032\pm0.297$ & $1.322\pm0.842$ & $1.084\pm0.100$\tabularnewline
\hline 
$\Omega_{b}^{-}\rightarrow\Xi^{*0}$ & $-0.048\pm0.016$ & $0.000\pm0.000$ & $0.046\pm0.015$ & $-0.145\pm0.049$ & $0.009\pm0.007$ & $0.019\pm0.007$ & $-0.050\pm0.022$ & $0.119\pm0.048$\tabularnewline
\hline 
$\Omega_{b}^{-}\rightarrow\Omega_{c}^{*0}$ & $-0.362\pm0.035$ & $0.017\pm0.006$ & $0.332\pm0.025$ & $-0.816\pm0.075$ & $-0.165\pm0.019$ & $0.006\pm0.003$ & $-0.034\pm0.047$ & $0.510\pm0.071$\tabularnewline
\hline 
\end{tabular}
}
\end{table}

\subsection{Semileptonic decays}

The vector and axial-vector helicity amplitudes are defined by
\begin{align}
H_{\lambda^{\prime},\lambda_{W}}^{V} & \equiv\langle\mathcal{B}_{f}^{*}(\lambda^{\prime})|\bar{q}\gamma^{\mu}Q|\mathcal{B}_{i}(\lambda)\rangle\epsilon_{W\mu}^{*}(\lambda_{W}),\nonumber \\
H_{\lambda^{\prime},\lambda_{W}}^{A} & \equiv\langle\mathcal{B}_{f}^{*}(\lambda^{\prime})|\bar{q}\gamma^{\mu}\gamma_{5}Q|\mathcal{B}_{i}(\lambda)\rangle\epsilon_{W\mu}^{*}(\lambda_{W}),
\end{align}
where $\lambda=\lambda_{W}-\lambda^{\prime}$. These helicity amplitudes
can be expressed in terms of the form factors:
\begin{align}
H_{3/2,1}^{V,A} & =\mp i\sqrt{2MM^{\prime}(\omega\mp1)}f_{4}^{V,A},\nonumber \\
H_{1/2,1}^{V,A} & =i\sqrt{\frac{2}{3}}\sqrt{MM^{\prime}(\omega\mp1)}\left[f_{4}^{V,A}-2(\omega\pm1)f_{1}^{V,A}\right],\nonumber \\
H_{1/2,0}^{V,A} & =\pm i\frac{1}{\sqrt{q^{2}}}\frac{2}{\sqrt{3}}\sqrt{MM^{\prime}(\omega\mp1)}\nonumber \\
 & \times\left[(M\omega-M^{\prime})f_{4}^{V,A}\mp(M\mp M^{\prime})(\omega\pm1)f_{1}^{V,A}+M^{\prime}(\omega^{2}-1)f_{2}^{V,A}+M(\omega^{2}-1)f_{3}^{V,A}\right],
\end{align}
where $\omega\equiv v\cdot v^{\prime}=P\cdot P^{\prime}/MM^{\prime}$,
and $f_{i}^{V,A}$ stand for the vector and axial-vector form factors,
respectively. The remaining helicity amplitudes can be obtained through
\begin{equation}
H_{-\lambda^{\prime},-\lambda_{W}}^{V,A}=\mp H_{\lambda^{\prime},\lambda_{W}}^{V,A},
\end{equation}
and the total helicity amplitudes are defined by 
\begin{equation}
H_{\lambda^{\prime},\lambda_{W}}=H_{\lambda^{\prime},\lambda_{W}}^{V}-H_{\lambda^{\prime},\lambda_{W}}^{A}.
\end{equation}
The transverse and longitudinal polarized differential decay widths
are 
\begin{align}
\frac{d\Gamma_{T}}{d\omega} & =\frac{G_{F}^{2}}{(2\pi)^{3}}|V_{\text{CKM}}|^{2}\frac{q^{2}M^{\prime2}\sqrt{\omega^{2}-1}}{12M}\left[|H_{1/2,1}|^{2}+|H_{-1/2,-1}|^{2}+|H_{3/2,1}|^{2}+|H_{-3/2,-1}|^{2}\right],\nonumber \\
\frac{d\Gamma_{L}}{d\omega} & =\frac{G_{F}^{2}}{(2\pi)^{3}}|V_{\text{CKM}}|^{2}\frac{q^{2}M^{\prime2}\sqrt{\omega^{2}-1}}{12M}\left[|H_{1/2,0}|^{2}+|H_{-1/2,0}|^{2}\right].
\end{align}

The numerical results for semileptonic decays are collected in Table
\ref{Tab:semi}. 

\begin{table}
\caption{Our predictions on semileptonic decays, where the uncertainties
from the form factors are considered.}
\label{Tab:semi} %
\begin{tabular}{c|c|c|c}
\hline 
Channel & $\Gamma/\mathrm{GeV}$ & ${\cal B}$ & $\Gamma_{L}/\Gamma_{T}$\tabularnewline
\hline 
$\Omega_{c}^{0}\rightarrow\Xi^{*-}e^{+}\nu_{e}$ & $(4.27\pm1.19)\times10^{-15}$ & $(1.77\pm0.49)\times10^{-3}$ & $0.84\pm0.01$\tabularnewline
\hline 
$\Omega_{c}^{0}\rightarrow\Omega^{-}e^{+}\nu_{e}$ & $(1.45\pm0.32)\times10^{-13}$ & $(6.03\pm1.34)\times10^{-2}$ & $1.02\pm0.00$\tabularnewline
\hline 
$\Omega_{b}^{-}\rightarrow\Xi^{*0}e^{-}\bar{\nu}_{e}$ & $(3.33\pm3.08)\times10^{-17}$ & $(8.29\pm7.69)\times10^{-5}$ & $0.92\pm0.00$\tabularnewline
\hline 
$\Omega_{b}^{-}\rightarrow\Omega_{c}^{*0}e^{-}\bar{\nu}_{e}$ & $(1.46\pm0.39)\times10^{-14}$ & $(3.64\pm0.97)\times10^{-2}$ & $1.26\pm0.02$\tabularnewline
\hline 
\end{tabular}
\end{table}

\subsection{Nonleptonic decays}

As mentioned earlier, for nonleptonic decays, we only consider the contribution of the external W-emission diagram.

When the meson in the final state is a pseudoscalar, the decay width
is \cite{Zhao:2018mrg}
\begin{equation}
\Gamma=|\lambda|^{2}f_{P}^{2}\frac{M|\vec{P}^{\prime}|^{3}}{6\pi M^{\prime}}\left[(\omega-1)(B^{2}-2AB)+2A^{2}\omega\right],
\end{equation}
where
\begin{align}
\lambda & \equiv\frac{G_{F}}{\sqrt{2}}\xi a_{1},\nonumber \\
A & =(M-M^{\prime})\frac{g_{1}}{M}+\frac{g_{2}}{M^{2}}(P\cdot q)+\frac{g_{3}}{MM^{\prime}}(P^{\prime}\cdot q)+g_{4,}\nonumber \\
B & =-(M+M^{\prime})\frac{f_{1}}{M}+\frac{f_{2}}{M^{2}}(P\cdot q)+\frac{f_{3}}{MM^{\prime}}(P^{\prime}\cdot q)+f_{4}.
\end{align}
with $\xi$ the relevant CKM matrix elements. When the meson in the
final state is a vector, the decay width is \cite{Zhao:2018mrg}
\begin{equation}
\Gamma=|\lambda|^{2}f_{V}^{2}m^{2}\frac{|\overrightarrow{P}^{\prime}|}{16\pi M^{2}}\left[|H_{1/2,1}|^{2}+|H_{-1/2,-1}|^{2}+|H_{3/2,1}|^{2}+|H_{-3/2,-1}|^{2}+|H_{1/2,0}|^{2}+|H_{-1/2,0}|^{2}\right],
\end{equation}
where $m$ is the meson mass.

The corresponding numerical results are collected in Table \ref{Tab:non}. 

\begin{table}
\caption{Our predictions on nonleptonic decays, where the uncertainties from
the form factors are considered.}
\label{Tab:non}%
\resizebox{\textwidth}{!}{%
\begin{tabular}{c|c|c|c|c|c}
\hline 
Channel & $\Gamma/\mathrm{GeV}$ & $\mathcal{B}$ & Channel & $\Gamma/\mathrm{GeV}$ & $\mathcal{B}$\tabularnewline
\hline 
$\Omega_{c}^{0}\rightarrow\Xi^{*-}\pi^{+}$ & $(1.70\pm0.48)\times10^{-15}$ & $(7.07\pm2.00)\times10^{-4}$ & $\Omega_{c}^{0}\rightarrow\Xi^{*-}\rho^{+}$ & $(1.14\pm0.32)\times10^{-14}$ & $(4.72\pm1.32)\times10^{-3}$\tabularnewline
\hline 
$\Omega_{c}^{0}\rightarrow\Xi^{*-}K^{+}$ & $(1.22\pm0.30)\times10^{-16}$ & $(5.04\pm1.24)\times10^{-5}$ & $\Omega_{c}^{0}\rightarrow\Xi^{*-}K^{*+}$ & $(6.46\pm1.79)\times10^{-16}$ & $(2.68\pm0.74)\times10^{-4}$\tabularnewline
\hline 
$\Omega_{c}^{0}\rightarrow\Omega^{-}\pi^{+}$ & $(1.11\pm0.27)\times10^{-13}$ & $(4.61\pm1.12)\times10^{-2}$ & $\Omega_{c}^{0}\rightarrow\Omega^{-}\rho^{+}$ & $(5.34\pm1.16)\times10^{-13}$ & $(2.21\pm0.48)\times10^{-1}$\tabularnewline
\hline 
$\Omega_{c}^{0}\rightarrow\Omega^{-}K^{+}$ & $(6.22\pm1.17)\times10^{-15}$ & $(2.58\pm0.49)\times10^{-3}$ & $\Omega_{c}^{0}\rightarrow\Omega^{-}K^{*+}$ & $(2.48\pm0.51)\times10^{-14}$ & $(1.03\pm0.21)\times10^{-2}$\tabularnewline
\hline 
$\Omega_{b}^{-}\rightarrow\Xi^{*0}\pi^{-}$ & $(5.89\pm4.85)\times10^{-19}$ & $(1.47\pm1.21)\times10^{-6}$ & $\Omega_{b}^{-}\rightarrow\Xi^{*0}\rho^{-}$ & $(1.41\pm1.18)\times10^{-18}$ & $(3.50\pm2.94)\times10^{-6}$\tabularnewline
\hline 
$\Omega_{b}^{-}\rightarrow\Xi^{*0}K^{-}$ & $(4.85\pm4.01)\times10^{-20}$ & $(1.21\pm1.00)\times10^{-7}$ & $\Omega_{b}^{-}\rightarrow\Xi^{*0}a_{1}^{-}$ & $(2.09\pm1.78)\times10^{-18}$ & $(5.20\pm4.45)\times10^{-6}$\tabularnewline
\hline 
$\Omega_{b}^{-}\rightarrow\Xi^{*0}D^{-}$ & $(1.07\pm0.93)\times10^{-19}$ & $(2.68\pm2.32)\times10^{-7}$ & $\Omega_{b}^{-}\rightarrow\Xi^{*0}K^{*-}$ & $(7.43\pm6.26)\times10^{-20}$ & $(1.85\pm1.56)\times10^{-7}$\tabularnewline
\hline 
$\Omega_{b}^{-}\rightarrow\Xi^{*0}D_{s}^{-}$ & $(2.93\pm2.55)\times10^{-18}$ & $(7.30\pm6.37)\times10^{-6}$ & $\Omega_{b}^{-}\rightarrow\Xi^{*0}D^{*-}$ & $(1.44\pm1.28)\times10^{-19}$ & $(3.60\pm3.18)\times10^{-7}$\tabularnewline
\hline 
- & - & - & $\Omega_{b}^{-}\rightarrow\Xi^{*0}D_{s}^{*-}$ & $(3.61\pm3.20)\times10^{-18}$ & $(8.99\pm7.97)\times10^{-6}$\tabularnewline
\hline 
$\Omega_{b}^{-}\rightarrow\Omega_{c}^{*0}\pi^{-}$ & $(1.31\pm0.30)\times10^{-15}$ & $(3.25\pm0.74)\times10^{-3}$ & $\Omega_{b}^{-}\rightarrow\Omega_{c}^{*0}\rho^{-}$ & $(3.41\pm0.80)\times10^{-15}$ & $(8.50\pm1.99)\times10^{-3}$\tabularnewline
\hline 
$\Omega_{b}^{-}\rightarrow\Omega_{c}^{*0}K^{-}$ & $(1.02\pm0.23)\times10^{-16}$ & $(2.55\pm0.58)\times10^{-4}$ & $\Omega_{b}^{-}\rightarrow\Omega_{c}^{*0}a_{1}^{-}$ & $(4.54\pm1.10)\times10^{-15}$ & $(1.13\pm0.27)\times10^{-2}$\tabularnewline
\hline 
$\Omega_{b}^{-}\rightarrow\Omega_{c}^{*0}D^{-}$ & $(1.05\pm0.26)\times10^{-16}$ & $(2.61\pm0.66)\times10^{-4}$ & $\Omega_{b}^{-}\rightarrow\Omega_{c}^{*0}K^{*-}$ & $(1.76\pm0.41)\times10^{-16}$ & $(4.38\pm1.03)\times10^{-4}$\tabularnewline
\hline 
$\Omega_{b}^{-}\rightarrow\Omega_{c}^{*0}D_{s}^{-}$ & $(2.59\pm0.66)\times10^{-15}$ & $(6.46\pm1.64)\times10^{-3}$ & $\Omega_{b}^{-}\rightarrow\Omega_{c}^{*0}D^{*-}$ & $(2.40\pm0.63)\times10^{-16}$ & $(5.99\pm1.56)\times10^{-4}$\tabularnewline
\hline 
- & - & - & $\Omega_{b}^{-}\rightarrow\Omega_{c}^{*0}D_{s}^{*-}$ & $(5.73\pm1.51)\times10^{-15}$ & $(1.43\pm0.38)\times10^{-2}$\tabularnewline
\hline 
\end{tabular}
}
\end{table}

\subsection{Comparison}

In this subsection, we compare our predictions with existing ones in the literature,
as can be seen in Table \ref{Tab:comparison}. Some comments are in order. 
\begin{itemize}
\item In Table \ref{Tab:comparison}, our predictions on the semileptonic
decay $\Omega_{b}^{-}\rightarrow\Omega_{c}^{*0}e^{-}\bar{\nu}_{e}$
are compared with those in Refs. \cite{Lu:2023rmq} and \cite{Ebert:2006rp}.
It can be seen that, our results are comparable to those in the literature.
It is worth noting that Ref. \cite{Lu:2023rmq} employs the same methodology
as this work, albeit with slightly different parameter choices, while
Ref. \cite{Ebert:2006rp} adopts a relativistic quark model under
the diquark picture.
\item In Table \ref{Tab:comparison}, our predictions on the nonleptonic
decays $\Omega_{c}^{0}\rightarrow\Omega^{-}\pi^{+}$ and $\Omega_{c}^{0}\rightarrow\Omega^{-}\rho^{+}$
are compared with those in the literature. One specific detail deserves
attention. Similar to this work, Refs. \cite{Wang:2022zja} and \cite{Cheng:1996cs}
provide decay widths, while other references provide decay branching
ratios. For the latter case, we need to pay attention to the lifetime of
$\Omega_{c}^{0}$. Specifically, in Refs. \cite{Shi:2025ocl}, \cite{Zeng:2024yiv},
and \cite{Hsiao:2020gtc}, $\tau_{\Omega_{c}}$ is taken as 268 fs,
while in Ref. \cite{Liu:2023dvg}, it is 273 fs. From Table \ref{Tab:comparison}, it can be seen that there are significant differences in the decay widths of $\Omega_{c}^{0}\rightarrow\Omega^{-}\pi^{+}$ and $\Omega_{c}^{0}\rightarrow\Omega^{-}\rho^{+}$ among different literature, and our results
are basically consistent with those in Refs. \cite{Zeng:2024yiv}
and \cite{Cheng:1996cs}, where the authors all adopted non-relativistic
quark models. 
\item In Table \ref{Tab:comparison}, our prediction on $\mathcal{R}_{e/\pi}$
is also compared with some other theoretical predictions and experimental
measurements, where $\mathcal{R}_{e/\pi}$ is defined as
\begin{equation}
\mathcal{R}_{e/\pi}=\frac{\mathcal{B}_{\Omega_{c}^{0}\rightarrow\Omega^{-}e^{+}\nu_{e}}}{\mathcal{B}_{\Omega_{c}^{0}\rightarrow\Omega^{-}\pi^{+}}}.
\end{equation}
It can be seen that, our result is consistent with most existing results
in the literature, but are somewhat smaller than Belle's data.
\end{itemize}

\begin{table}
\caption{Our predictions are compared with those in the literature.}
\label{Tab:comparison}
\begin{tabular}{c|c|c|c}
\hline 
$\Omega_{b}^{-}\rightarrow\Omega_{c}^{*0}e^{-}\bar{\nu}_{e}$ & This work & Lu \cite{Lu:2023rmq} & Ebert \cite{Ebert:2006rp}\tabularnewline
\hline 
$\Gamma/10^{-14}\,{\rm GeV}$ & $1.46\pm0.39$ & $1.31$ & $1.99$\tabularnewline
\hline 
$\Gamma_{L}/\Gamma_{T}$ & $1.26\pm0.02$ & $1.18$ & $0.95$\tabularnewline
\hline 
\end{tabular}
\begin{tabular}{c|c|c|c|c|c|c|c}
\hline 
 & This work & Shi \cite{Shi:2025ocl} & Zeng \cite{Zeng:2024yiv} & Liu \cite{Liu:2023dvg} & Wang \cite{Wang:2022zja} & Hsiao \cite{Hsiao:2020gtc} & Cheng \cite{Cheng:1996cs}\tabularnewline
\hline 
$\Gamma(\Omega_{c}^{0}\rightarrow\Omega^{-}\pi^{+})/10^{-14}\,{\rm GeV}$ & $11.1\pm2.7$ & $0.37\pm0.20$ & $8.42\pm1.18$ & $4.53\pm0.36$ & $2.6$ & $1.25\pm0.17$ & $10.6$\tabularnewline
\hline 
$\Gamma(\Omega_{c}^{0}\rightarrow\Omega^{-}\rho^{+})/10^{-14}\,{\rm GeV}$ & $53.4\pm11.6$ & $17.9\pm5.2$ & $44.9\pm4.8$ & - & - & $3.53\pm0.10$ & $37.3$\tabularnewline
\hline 
\end{tabular}
\begin{tabular}{c|c|c|c|c|c|c}
\hline 
 & This work & Zeng \cite{Zeng:2024yiv} & Hsiao \cite{Hsiao:2020gtc} & Alice \cite{ALICE:2024xjt} & Belle \cite{Belle:2021dgc} & CLEO \cite{CLEO:2002imi}\tabularnewline
\hline 
$\mathcal{R}_{e/\pi}$ & $1.31\pm0.02$ & $1.18\pm0.22$ & $1.1\pm0.2$ & $1.12\pm0.35$ & $1.98\pm0.15$ & $2.4\pm1.2$\tabularnewline
\hline 
\end{tabular}
\end{table}

\section{Conclusions}

This work is devoted to investigating the $1/2\to3/2$ weak decays
of singly heavy baryons. Due to the orthogonality between the spin
wavefunctions of antitriplet baryons and spin-3/2 baryons, the weak
decay amplitude for such processes vanishes at the leading order of QCD. Consequently,
this study exclusively examines the weak decays of $\Omega_{Q}$ --
the only member of the singly heavy baryon sextet that predominantly
decays via weak interactions.

Within the framework of the light-front quark model under the three-quark
picture, we systematically study the $1/2\to3/2$ weak transitions
of $\Omega_{Q}$. First, we extract the relevant form factors using
this approach. These form factors are then applied to investigate
corresponding semileptonic and nonleptonic decay channels. Finally,
we compare our phenomenological predictions with existing results
in the literature. Our findings are expected to be helpful in experimentally
establishing these decay channels.

\section*{Acknowledgements}

The authors are grateful to Prof. Zhi-Gang Wang for valuable discussions.
This work is supported in part by National Natural Science Foundation
of China under Grant No.~12465018.

\end{document}